\documentclass{elsart}

\usepackage[latin1]{inputenc}
\usepackage{amsmath}
\usepackage{amsfonts}
\usepackage{amssymb}
\usepackage{graphicx}

\begin{document}

\begin{frontmatter}
\title{A (reactive) lattice-gas approach to economic cycles }

\author[liege]{Marcel Ausloos},
\ead{Marcel.Ausloos@ulg.ac.be}
\author[grasp]{Paulette Clippe},
\ead{P.Clippe@ulg.ac.be}
\author[liege,ar]{Janusz Mi\'skiewicz} 
\ead{jamis@ozi.ar.wroc.pl} and 
\author[ift]{Andrzej Pekalski}
\ead{apekal@ift.uni.wroc.pl}

\address[liege]{SUPRAS (a member of SUPRATECS) and GRASP, B5, Sart Tilman Campus B-4000 Li$\grave e$ge, Euroland }

\address[grasp]{GRASP, B5, University of Li$\grave e$ge, B-4000 Li$\grave e$ge, Euroland }

\address[ar]{Department of Physics and Biophysics, University of Agriculture, ul. Norwida 25, 50-375 Wroc{\l}aw, Poland}

\address[ift]{Institute of Theoretical Physics, University of Wroc{\l}aw, pl. Maxa Borna 9, PL 50-204 Wroc{\l}aw, Poland}

\begin{abstract}

A microscopic approach to  macroeconomic  features is intended. A model for
macroeconomic behavior under heterogeneous spatial economic conditions is
reviewed. A birth-death lattice gas model taking into account the 
influence of an
economic environment on the  fitness and concentration evolution of economic
entities is numerically and analytically examined. The reaction-diffusion model
can be also mapped onto a high order logistic map. The role of the selection
pressure along various dynamics with entity diffusion on a square symmetry
lattice has been studied by Monte-Carlo simulation. The model leads 
to a sort of
phase transition for the fitness gap as a function of the selection 
pressure and
to cycles. The control parameter is a (scalar) ''business plan''. The business
plan(s) allows for spin-offs or merging and enterprise survival 
evolution law(s),
whence bifurcations, cycles  and chaotic behavior.
\end{abstract} 
\begin{keyword}
econophysics, macroeconomy, evolution, external field, selection
\PACS{89.65.Gh, 05.10.Ln,  89.75.-k, 07.05.Tp, 05.65.+b}
\end{keyword}
\end{frontmatter}

\maketitle

\section{Introduction}

Econophysics \cite{Takayasu0,Takayasu1,Takayasu2,MantegnaStanleybook,BouchaudPottersbook}
has been centered mainly about developing continuous models, writing time
dependent equations to describe fluxes, prices, assets, returns, risk, .... in
order to obtain evolutions or stable states \cite{Alchian,Boulding}. This leads
to Langevin-like, and/or Fokker-Planck equations  for price
evolutions 
\cite{friedrich1,AusloosIvanovaSP500Tsallis,russianapfa4,BouchaudContLange},
even e.g. Black-Scholes equation for options 
\cite{BouchaudPottersbook,Haven2002}
or simply partial distribution functions
description \cite{MantegnaStanleybook,AusloosIvanovaSP500Tsallis}. This is line
with Hamilton equation mechanistic approach which a classically trained
physicist can accept as a basic approach! Quantum mechanistic ideas are
sometimes coming in \cite{Haven2002}.

Another aspect has been to develop discrete, so called agent based, 
(microscopic)
models\footnote{an exhaustive list of references should be too long; see  for
some interesting review Samanidou et al. \cite{Samanidou}}, on 
lattices or not, on
networks or not, in order to  manipulate interactions and constraints 
in view of
describing such characteristics as prices, returns, etc. and perform 
simulations.
This is supposed to be interesting, according to economists, if the 
irrational or
not behavior of agents is  described as expected \cite{Wroclecon}, even though
physicists know that deterministic chaos exists. This 
''microscopic'' approach is
often coupled to data analysis, going from distribution functions to scaling
exponents and universal laws for various features, including
crashes \cite{Sornettebook}.

In the MACRO economy, theoretical work on  economic behaviors can 
also be thought
of to relate to some ''self-organized'' features 
\cite{Sornettebook,Bak}. In the
continuous time spirit, the variables of interest are sometimes GDP,
immigration rate, wages, price, wealth, profit, taxes, capital growth,... The
equations again are of the Langevin type and contain coupling parameters which
are hardly measurable \cite{Solvay,Gligor}, and are mostly ''ad 
hoc'', even though
based on so called ''financial theory'' expectations. One question 
exists whether
for either rarely well quantified or abundantly quantified, in a statistical
sense, many macroeconomic features  can be described from a microscopic-like
level, with understandable parameters \cite{BerghGowdy01}. This is in line with
previous  attempt to connect macroeconomy and econophysics \cite{reconcile}.

 From a macroeconomic behavior point of view, it is remarkable that there are
spatio-temporal changing economic conditions, - and a variable space as well.
E.g. after the Berlin wall opening, there was a sort of  "physical volume, or
available space" increase  for economic entities, like  for particles in a
container. The gas particles are hereby called companies, but this is 
only a name
for some economic variables. The company ''degree of
freedom'' (particle efficiency $f$) can be coupled to an external field $F$.
Beside the field, there is some selection pressure $sel$. Moreover as for a gas,  economic
entities are allowed to move. Also the $f$ value of the new firms was 
considered
to be obtained according to various types of memories depending on 
the $f$ of the
company parents \cite{ACPbali}. The algorithm of our toy-like
model \cite{ACPbali}, a chemically reactive lattice gas, is recalled 
in Sect.2. It
was aimed at investigating whether macroeconomic features, like so called
economic 
cycles\footnote{$http://cepa.newschool.edu/het/schools/business.htm$} 
\cite{cyclebook1,cyclebook2} could be obtained
or recovered from elementary rules with (''microscopic'') interactions between
entities characterized by some degree of freedom (their efficiency) 
coupled to a
field which can be qualitatively considered an economic one.  A so called
business plan, based on a merging and spin-off creation alternative was
considered for the enterprise concentration evolution. An {\it a priori law} is
given for the efficiency evolution.

In section 3, we outline a few results, like
the concentration and efficiency of entity evolutions, comparing some 
stochastic or not
initiating conditions.  From Monte-Carlo simulations it is  observed 
that the model
leads to a phase
transition or self-organisation-like scenario as a function of the selection
pressure.  Some analytical results are also presented within a mean 
field approximation
for the best adapted company under constant field conditions. A short
conclusion is  found in Sect. 4.

\section{Model and algorithm sketch}

Initially all firms are located at random positions on a lattice and receive a
random $f$ value. The algorithm, based on essentially three rules 
($R_i$) is:

\begin{enumerate} 
\item a firm ($i$)  is picked;
\item ($R_1$) a survival
probability

\begin{equation}
 p_i \,= \,\exp(-sel|f_i - F|) , 
\end{equation}

is checked against a  uniformly distributed random number $r_i \in [0,1]$. If
$r_i > p_i$  the firm is removed from the system;

\item ($R_2$) if the firm survived, then  the firm is moved to a site of the
nearest neighborhood, - if an empty place is found;

\item a random search is made  for a partner in the nearest neighborhood of the
new position. If found at the site $j$ then ($R_3$)

\item either  (with a probability $b$) the two firms merge, creating 
a new firm at the
location of the first one, with a new fitness $f_i$ while the second firm is
eliminated;

\item or (with a probability $1-b$) the two firms produce  (a 
predetermined amount of)
new ($k$) firms (spin-off's) with a (set of) fitness  $f_k$  in the Moore
neighborhood (9 sites on a square symmetry lattice) of the first firm.

\end{enumerate}

The three rules ($R_i$) of this  model are the minimalistic set of 
rules that we
can think of and yet yield a wealth of interesting evolutionary 
implications for
the economy. Of course, there are many different variants at each 
stage. E.g. to
pick up the firms, we have considered two cases: (i) either the pick is random
($P_1$), or (ii) the less adapted firm with respect to $F$ is chosen ($P_2$).

Many variants can be imagined. One could allow for a predetermined or 
not number
of searches for a move.  The search about a free nearest neighbor site could be
stochastic or deterministic. We only looked at a finite number of stochastic
searches.

One could give or not new values of the fitness to the nearest neighbors. These
values could be arbitrary, like in Bak-Sneppen original 
work \cite{BS}, or might
not be random. In our published work,  the lattice was always with square
symmetry, and the fitness evolution has always been driven by

\begin{equation} f_i (t+1) = \frac{1}{2}\left[(f_i (t) + f_j(t)) + 
sign[0.5 - r] |f_i(t) - f_j(t)|\right], 
\end{equation} 
where  $r$  is a random number in
[0,1].

The field, selection pressure, $b$ can change in space and time; feedback or
coupling between these parameters can be imposed; various boundary 
conditions can
be imposed... N.B. The time is measured in Monte-Carlo steps  (MCS). 
To complete
a  MCS one has to pick as many firms as there were at the beginning 
of that step.

\section{A few results}

\subsection{Simulations}

Results pertaining to two published cases \cite{ACPbali,ACP3BS} can 
be summarized;
all pertain to an {\it a priori b=0.01}. Many facets  can be revealed: the
strength of the selection pressure is primordial, for reaching 
asymptotic values,
but there are relatively well marked effects like

\begin {enumerate} 
\item some equilibrium between births and deaths;

\item the concentration $c_t$ can be maximal ($P_1$) or reach a finite value
($P_2$) (because in the latter,  more Darwinian case, case  the best adapted
companies can never die);

\item  the regions  where field gradients exist are prone to instabilities at
least \cite{ACP3BS} in case $P_2$;

\item the   economic  (''external'') field  implies stable  or unstable density
distributions (whence ''cycles'') ;

\item the diffusion process rule(s) are useful for invasion process, 
but are also
relevant for replinishing abandoned regions (whence ''cycles'');

\item  the average fitness is  more or less slowly reached  according to the
selection pressure;

\item  the   ''critical selection pressure''  depends on the dynamics chosen

\item the business plan effect is very complex (see next section for analytical
work about this).

\end{enumerate}

\subsection{Analytical approach}

In fact, the possible events can be easily enumerated, since two kinds of
birth-events, with respectively $b$ and $1-b$ probability, can be 
distinguished:
(i) the  number of $f$-states  does not change: in this case the amount of
companies existing on the net in the next MCS can be calculated by
multiplying the existing number of companies in a given state by the 
appropriate
survival probability; (ii) there is a change of the company state distribution.
This is the case of company merging or spin-off creation.  Considering all
possible events, in a ''mean field approximation'', the evolution 
equation of the distribution function $ N(t,f) $ can  be written as the sum of two terms

\begin{equation} 
\label{evol1} N (t+1,f) = H_1(c_t) p(f) N (t,f) + H_2(c_t) N
(t,g(f)) , 
\end{equation}

where the $H_1(c_t)$ and $H_2(c_t)$ polynomials (in $c_t$) containing 
coefficients
depending on the various possible processes can be found
elsewhere \cite{JMMALadek}. These polynomials mainly depend on the 
lattice symmetry
and the number of spin-off's which are created. The function $g(f)$ describes
the influence of the merging process on the distribution function $ N(t,f) $. 
In the case of best adapted companies \cite{JMMALadek} ($f=F$), the evolution equation for 
$N(t,g(F))$ can be simplified and written as a logistic equation of high order. Notice 
that in this case the $sel$ parameter is irrelevant.

\begin{figure}
\begin{center}
\includegraphics[scale=0.38, angle=-90]{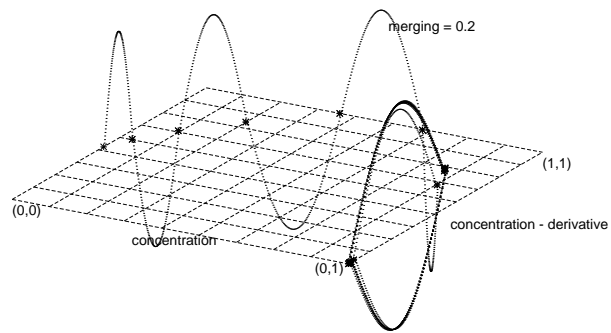} 
\includegraphics[scale=0.38, angle=-90]{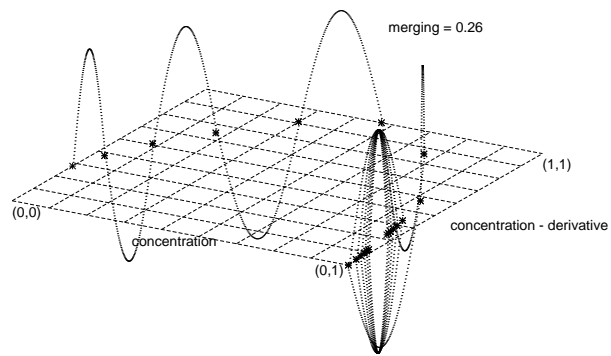} 
\includegraphics[scale=0.38, angle=-90]{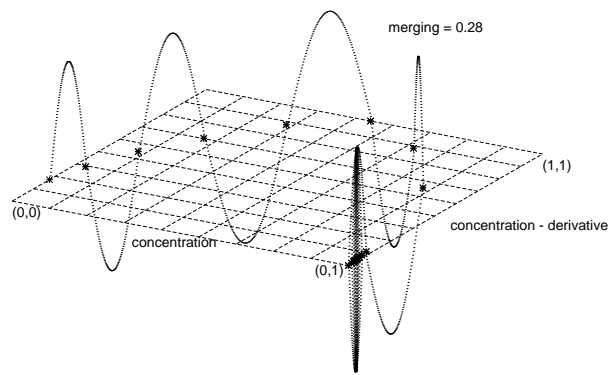} 
\end{center}
\caption{\label{fig_phase_20} Time evolution in parameter space ($c_t$, $\dot{c}_{t})$ of the
$N(t,g(F))$ trajectory in the intermediary $b$ regime characterized by cycles.}
\end{figure} 

In so doing we can focus on the $b$ parameter effect. It can be shown that the
system can (i) reach a  one stable solution   for $ b>0.45 $; (ii) oscillate
(Fig.1) with some characteristic time; (iii) display chaotic features  for   $
b<0.15 $, including as usual ''stability'' windows. Especially 
interesting is the
range $ b \in [0.38;0,45] $: damping properties are superposed to an 
oscillating
behavior as seen from a study of a generalized Lyapunov
exponent \cite{JMMALadek}.

\section {Conclusions}

Changes in related {\it microeconomic} conditions may induce a change in the
efficiency of a  {\it macroeconomy}.  For instance, a modification of 
traffic laws may
distort the economic efficiency of a company. Likewise, social or cultural
changes, perhaps driven by technological innovations, may induce an economic
modification without any changes to the basic economic conditions 
themselves. The
most important aspect of the above is to recognize that one does not need to
stick to continuity evolution equations in order to describe such a 
macroeconomy
evolution. We  have thus presented a  ''death  and birth reactive lattice gas
process'' along a microscopic physics like approach  in order to describe  a
specific macroeconomy evolution. No need to say that the behavior, of a
macroeconomy  is of much greater complexity than as done here above. 
Fortunately many improvements are possible and needed.

\vskip 0.6cm

{\bf Acknowledgments}

\vskip 0.6cm

MA and AP thank the CGRI and KBN for partial financial support allowing mutual
visits during this work. MA and PC thank  an Action Concert\'ee Program of the
University of Li$\grave e$ge (ARC 02/07-293).  JM thanks FNRS for financial
support and GRASP for the welcome and hospitality.

\end{document}